# Optimized FTO seeding enables the growth of high efficient Ta-doped TiO₂ nanorod photoanodes

Christopher Schneider,[a] Ning Liu,[a] Patrik Schmuki*[a]

[a]*Department of Materials Science, WW4-LKO, University of Erlangen-Nuremberg, Martensstrasse 7, D-91058 Erlangen, Germany. E-mail: Schmuki@ww.uni-erlangen.de*

[b]*Department of Chemistry, King Abdulaziz University, Jeddah, Saudi Arabia*

Link to the published article:
https://pubs.rsc.org/en/content/articlehtml/2017/cc/c7cc05168a

**Tantalum doped rutile nanorods were hydrothermally grown on FTO substrates using a new seeding approach. This approach allows the incorporation of high concentrations of up to 4.8 at% tantalum as active doping and results in a significant enhancement of photoelectrochemical water splitting rate (1.8 mA/cm² at a potential of +1.5 V vs RHE) which corresponds to ~ 1% photocurrent conversion efficiency under AM 1.5, 100 mW/cm² simulated sunlight irradiation.**

Since the first reports by Fujishima and Honda [1] titanium oxide is well known for its ability to serve as a semiconductor for the photoelectrolysis of water into $H_2$ and $O_2$. In spite of inherent drawbacks (large band-gap of 3.0 eV that allows only UV light absorbtion), $TiO_2$ is still considered as one of the most promising candidates as a photoanode for water splitting, due to its strong optical UV absorption, high chemical stability and low cost [2] . One-dimensional morphologies of $TiO_2$ such as nanotubes or nanorods offer direct electrical pathways for photogenerated electrons which is considered a key beneficial feature for an enhancement of the performance of photoelectrochemical devices. Lately, hydrothermally grown rutile nanorods have attracted much attention because of their comparably high photoconversion efficiencies due to their single crystalline and 1D nature. These nanorods are typically grown on FTO, as the small lattice mismatch of 2 % between FTO and rutile crystal structure enables the direct hydrothermal growth onto this substrate [3]. To increase efficiency of the nanorod arrays, several approaches were studied in the past few years, which include hydrothermally grown 3D nanostructures, decoration with graphene quantum dots, building nanocomposites or more advanced heterostructures with rutile/anatase phase junctions.[4] Still one major drawback for these photoanodes lies in the fact that rutile has a comparable low electron mobility [5]. To overcome this issue, various methods have been developed, such as post synthesis hydrogen reduction [6] or doping of nanorods and nanotubes with various metal and non-metals [7]. Doping allows to increase the rutile conductivity by doping with suitable donors, such as small amounts of tantalum [8]. Over the past years various efforts targeted the doping of rutile rods, namely with tantalum, but the resulting photocurrents in water splitting experiments remained quite low. However, as we will show in this work, a key factor for successful doping is the seeding process and the nature of the seeds from which tantalum doped nanorods are grown.

$TiO_2$ nanoparticle seeding is often used to provide initiation sites with a defined density. After a uniform hydrothermal growth, such nanorods are typically adherent to the FTO surface [8]. It has also been shown that rutile nanorods can be grown onto other substrates, such as silicon, glass and titanium metal by treating the surface with $TiO_2$ nanoparticles and $TiO_2$ thin films [9]. The resulting structures can be influenced by the titanium precursor concentration or by adjusting synthesis conditions [3]. Nevertheless these conventional seeding procedures fail, if the direct growth of doped $TiO_2$ rutile nanorods is attempted (as we will show below). The present work introduces a seeding strategy on FTO that not only enables the successful growth tantalum doped nanorods as shown in Figure 1a, but provides structures with a greatly enhanced performance in photoelectochemical water splitting.

In a first set of experiments that targeted a controlled growth of tantalum doped rutile nanorods, we tried a titanium nanoparticle seeded approach (for experimental details see SI). In this approach a $TiCl_4$ treatment was used to produce anatase nanoparticles on a substrate [10], onto which nanorods can then be hydrothermally grown. These experiments showed that, seeding of the FTO substrate was mandatory, if a direct tantalum doping was attempted via a mixed titanium/tantalum precursor in the hydrothermal experiments. In fact the addition of the tantalum precursor hindered the growth of the rutile nanorods to such an extend it was not possible to grow tantalum doped nanorods at all on a bare FTO substrate. When this conventional $TiO_2$ nanoparticle seeding (Figure S1a), which enabled the growth of nanorods from titanium-tantalum mixed precursor solutions was used, the resulting rod arrays showed a drastic decrease in water splitting activity even when compared to undoped narorods (Figure 1d).

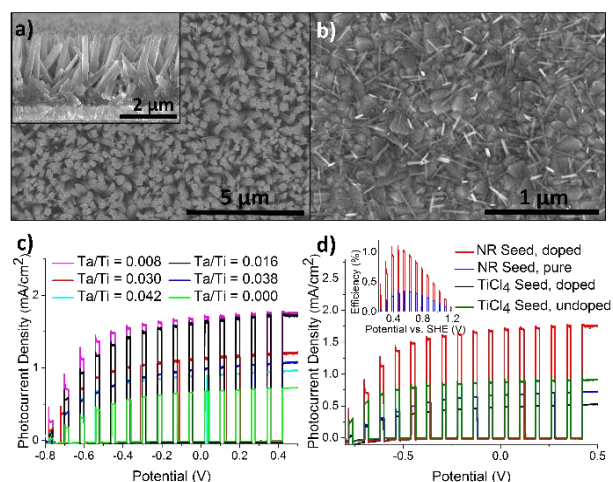

Figure 1 a) SEM image of vertically aligned tantalum doped rutile nanorods and cross section of tantalum doped rutile nanorods (upper left) b) SEM image of a rutile nanorod seeded FTO substrate, c) PEC water splitting curves for Ta-doped rutile nanorods with different doping ratios grown onto Ru-Nr seeded FTO annealed at 450 °C d) PEC water splitting curves for pure and tantalum doped rutile nanorods grown on rutile nanorod and $TiO_2$ nanoparticle seeded samples respectively, annealed at 450 °C and photoconversion efficiencies calculated for pure and doped samples with NR seeding (upper left)

Therefore we explored alternate seeding strategies for direct hydrothermal tantalum doped nanorod growth. The by far most successful attempt turned out to be the use of small undoped rutile nanorods. Figure 1b) shows an SEM image of an FTO surface with these non-doped seeding rods. For the seeding process, the FTO substrate was immersed in a titanium precursor solution with lower concentration and treated at a lower temperature of 150 °C for 4h. From these seeds, the tantalum doped nanorods can start growing successfully into single crystalline tantalum doped rutile nanorods as shown in Figure 1a. To investigate the influence of the doping process on the PEC activity of the rods, PEC water splitting experiments were conducted, as outlined in the experimental section, in 1 M KOH using AM 1.5 conditions at 100mW/cm$^2$. To screen for an optimized doping concentration, we varied the ratio of Ta/Ti precursor. In Figure 1c the maximum photocurrent for various doping ratios is shown. Clearly at a Ta/Ti precursor ratio of 0.008 a maximum photocurrent is observed. With higher doping ratios, the maximum photocurrent decreases, due to the increasing numbers of recombination centers. For further characterization, we used the best performing Ta/Ti = 0.008 sample. The I-V characteristics of these optimally performing rutile nanorod seeded, tantalum doped nanorods reach a maximum photocurrent of ~1.8 mA/cm$^2$ – this is far superior to the non-doped TiO$_2$ rods that only reach around 0.8 mA/cm$_2$ (the TiCl$_4$ seeded tantalum doped samples perform even worse). If the tantalum content in a TiCl$_4$ seeded sample and a rutile seeded sample, grown under identical hydrothermal conditions are compared, one finds over 8 times more tantalum for the rutile seeded samples (calculated from Figure 2e and Figure S2b).

To evaluate the efficiency of PEC hydrogen generation from doped and undoped rutile nanorod samples, the photoconversion efficiency as shown in Figure 1d) was calculated based on the equation, η =I (1.23 – V)/J [11]. I is the photocurrent density at the measured potential, V is the applied voltage versus the reversible hydrogen electrode (RHE) and J is the irradiance intensity of 100 mW/cm$^2$ (AM 1.5). The reversible hydrogen electrode (RHE) potential can be converted from the Ag/AgCl reference electrode via the Nernst equation [6,12]. Considering the values for the used (3M KCl) Ag/AgCl reference and the 1M KOH solution this converts to: RHE = V(Ag/AgCl) + 1.04 V. The best sample with the initial volume ratio of V(Ta) / V(Ti) = 0.008 achieved an efficiency of ~1.0% at a bias of −0.52 V versus Ag/AgCl (0.52 V vs RHE), whereas the undoped sample only achieved ~0.33% at a voltage of -0.37 V versus Ag/AgCl ( 0.67 V vs RHE). This present a 200% increase and shows that tantalum doping can significantly increase the maximum photocurrent at lower potentials, which results in a highly improved photocurrent conversion efficiency.

A detailed analysis of the best performing rods used in our investigations with TEM, XPS and EDX are shown in Figure 2. XRD and SAD pattern only show rutile phase to be present and no significant tantalum phase could be detected. There was no significant change in the lattice found after tantalum incorporation as expected due to the comparable ionic radii (with VI coordination) of Ti$^{4+}$ (0.061 nm) and Ta$^{5+}$ (0.064 nm) ions.[13] As can be seen in figure 2f and g, the interplanar d-spacing of (110) planes for the tantalum doped nanorods determined from HRTEM is 0.3155 nm, compared to 0.3167 nm for the undoped TiO$_2$ nanorods. While the incorporation of tantalum into anatase has some minor effect on XRD spectra [14], no changes in the XRD could be observed in case of tantalum doped rutile [8,15]. From the TEM shown in Figure 2g, one can see that the (110) plane is parallel to the growth direction of the nanorods as previously reported [3] and this is also reflected in the TEM of the non-doped reference sample in Figure 2f). Although others have reported the appearance of a core-shell structure at doping concentrations of 2.6 atomic percent in a solvothermal method [8], we did not observe such structures. XPS spectra shown in Figure 2a, b, and c show the Ta4f peaks of Ta$_2$O$_5$, the Ti2p peaks and the O1s peaks respectively. The Ta/Ti ration according to XPS is 1.5% The EDX measurements seen in Figure 2e show an average tantalum concentration of 3.2 atomic percent which converts to a tantalum concentration of 4.8 atomic percent when corrected for the substrate tin and oxygen content. In the content of the crystal composition it is noteworthy that the average content, measured by EDX is significantly higher than the concentration in solution and on the surface, measured by XPS. This indicates the presence of a concentration gradient (with higher tantalum content at the bottom of the rods). A likely explanation is a difference in decomposition rate of the tantalum and

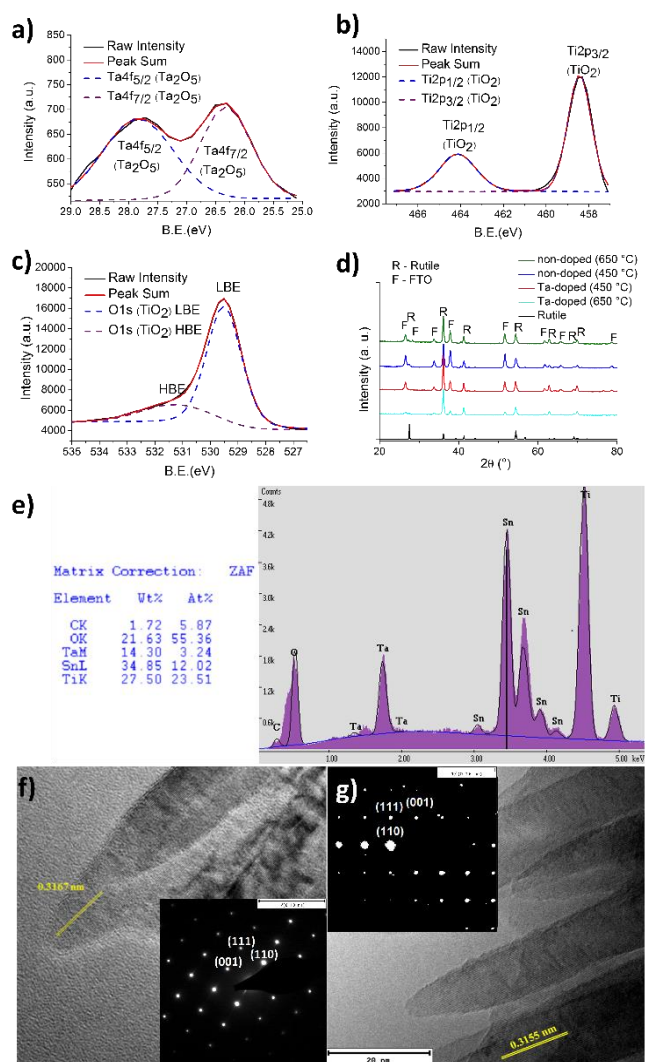

Figure 2 a)-c) XPS spectrum and d) XRD of tantalum doped and undoped TiO$_2$ nanorods for precursor ratios of Ti/Ta = 0.008 for different annealing temperatures e) EDX spectrum and concentration of elements of tantalum doped rutile nanorods according to EDX measurements for V(Ta) / V(Ti) = 0.008 f) TEM image and SAD pattern of non-doped and g) tantalum doped rutile nanorods the titanium precursor under hydrothermal conditions. At the end of the synthesis the tantalum precursor is mainly decomposed leading to a decreased tantalum to titanium ration in solution and near the surface. This may also explain the different tantalum concentrations in the differently seeded samples. It is reasonable to assume that the growth speed of the rutile nanorods is different on different surfaces, since in case of the rutile nanowire seeded sample there is no need for the nucleation of small rutile particles and rutile rods can just grow onto the already present rutile wires. In case of the nanoparticle seeded samples, there has to be nucleation of rutile particles or a transformation of the anatase particles to the rutile crystal structure which has been shown to occure under these reaction conditions [9]. The difference in both, growth and decomposition rate therefore leads to different tantalum concentration in the resulting rods, therefore changing their performance. Different decomposition rates thus can lead to gradients of tantalum concentration in the nanorods. To investigate this, Tof-SIMS sputter depth profiles, as shown in Figure 3a, were acquired. Figure 3b shows the Tof-SIMS spectra, which confirms that tantalum is present in the sample. The sputter depth profiles of Figure 3a show that in all cases, there is an increase in tantalum concentration with sputter depth, in line with above discussion and gradient formation.

Another factor that has been reported to affect the properties of tantalum doped TiO$_2$ is the annealing temperature [8]. On the one hand, an increase of the annealing temperature can reduce the activity of the nanorods due to thermal or mechanical stress at the rod/FTO interface or an increased sheet resistance of FTO[16], while on the other hand, a study on the desorption of chloride ions on hydrothermally grown rutile nanorods showed that a minimum annealing temperature is needed to remove the adsorbed chloride ions that block oxygen vacancies [17], which have been widely reported to be the main active sites for the dissociation of water and are therefore a significant reason for a hampered water splitting activity [18]. Since the concentration of the dopands and the annealing temperature strongly influence the effectiveness of the resulting material, differently doped nanorods were annealed at temperatures between 450 and 650 °C and investigated by PEC water splitting experiments conducted under simulated sunlight. As shown in Figure 3c, increasing the annealing temperature decreases the PEC water splitting activity significantly, while the XRD shows no change to crystal structure due to the increased temperature (Figure 2d). The maximum photocurrent for each doping ratio could be increased significantly by annealing at a temperature of 450 °C. This always resulted in the best performing nanorods, as can be seen in Figure S1c, with a more than 100% increase in water splitting activity compared to the undoped sample annealed at the same temperature and are close to the best reported values of doped rutile nanorods [7].

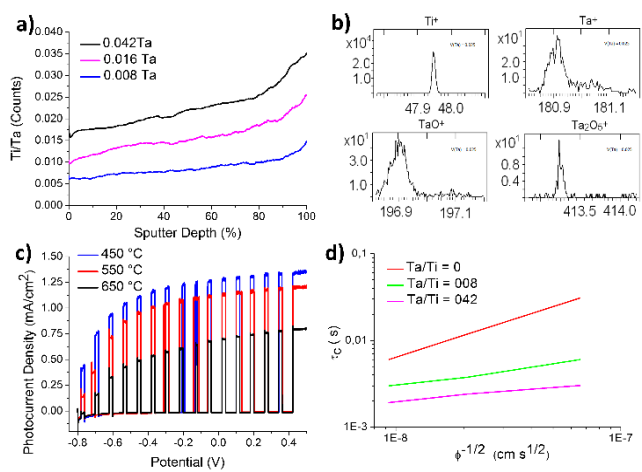

Figure 3 a) Tof-SIMS sputter depth profile of tantalum doped nanorods b) Tof-SIMS of tantalum doped rutile nanorods with starting precursor concentration of V(Ta) / V(Ti) = 0.016 c) PEC water splitting curves for tantalum doped rutile nanorods with a tantalum to titanium precursor volume ratio of 0.028, annealed at different temperatures nanorods d) Log-log graph of transport time constants calculated from IMPS measurements for doped and undoped nanowires annealed at 450°C vs Φ$^{-1/2}$.

To investigate the reasons for this improvement, photocurrent measurements, impedance spectroscopy, Mott-Schottky measurements and UV impedance spectroscopy were used to provide information on the spectral behavior as well as on the electronic properties of the samples as described in the supporting information. As can be seen in Figure S2a no red shift of the IPCE is seen in the 0.008 doped samples. The improvement of the water splitting activity results mainly from a generally increased IPCE and, in contradiction to previous reports [8], an increase of IPCE in the lower wavelengths region of the UV light which is likely caused by the increase of the electron diffusion length in tantalum doped rutile. From UV impedance spectra, the electron transport time constants T$_c$ were calculated. In Figure 3d a log–log graph of the transport time constant over Φ$^{-1/2}$ is shown for undoped and doped samples. The linear dependence of the transport time constant on the inverse square root of the photon flux in the logarithmic scale is in agreement with a mechanism driven by diffusion with T$_c$ reducing as light intensity increases. When comparing the doped and undoped sample, a 5 to 10 times higher transport time constant for the undoped sample is observed, which could explain the improved performance of the doped samples. From Mott-Schottky type measurements, the $N_D$ (the apparent donor density per cubic centimeters) was calculated as $9.11 \times 10^{19}$, $5.59 \times 10^{20}$ and $3.44 \times 10^{21}$ for the undoped, best and highly doped samples respectively as described in the supporting information (Figure S3).

In summary, the presented work shows that tantalum doped nanowires can be grown on rutile nanorods pre-seeded FTO substrates by a new seeding approach, which doubles the water splitting activity of the afterwards grown tantalum doped rutile nanorods. The efficiency of hydrothermally grown rutile nanorods decreases when annealed at temperature above 450 °C and depends strongly on dopant concentration. For optimized tantalum doped rods a maximum photocurrent of ~1.8 mA/cm$^2$ could be achieved, which is higher than for any previously reported tantalum doped rutile nanorods and up to 4 times higher than for undoped nanorod reference samples. This enhancement by tantalum doping is due to the improved electron transfer time constant and increased ICPE in the UV region of the light due to a larger carrier diffusion length. It is reasonable to predict that the maximum photocurrent and photo conversion efficiency can still be increased by post synthetic treatments like ammonium and hydrogen annealing. Therefore hydrothermally grown tantalum doped rutile nanowires have a high potential in any photoeletrochemical TiO$_2$ application.